\documentstyle[epsfig,rotate,preprint,floats,tighten,aps,amssymb]{revtex}

\def\bfgreek#1{ \mbox{\boldmath$#1$}}

\begin{document}
\draft

\preprint{\vbox{\noindent\null \hfill ADP-99-25/T365\\ 
                         \null \hfill hep-lat/9906027 \\
}}

\title{\huge Baryon Masses from Lattice QCD:\\
Beyond the Perturbative Chiral Regime}

\author{Derek B. Leinweber\footnote{dleinweb@physics.adelaide.edu.au}, 
        Anthony W. Thomas\footnote{athomas@physics.adelaide.edu.au}, 
        Kazuo Tsushima\footnote{ktsushim@physics.adelaide.edu.au},  and 
        Stewart V. Wright\footnote{swright@physics.adelaide.edu.au}}

\address{Department of Physics and Mathematical Physics\break
         and
         Special Research Centre for the Subatomic Structure of Matter,\break
         University of Adelaide, Australia 5005}
\maketitle

\begin{abstract}
Consideration of the analytic properties of pion-induced baryon
self-energies leads to new functional forms for the extrapolation of
light baryon masses.  These functional forms reproduce the leading
non-analytic behavior of chiral perturbation theory, the correct
non-analytic behavior at the $N \pi$ threshold and the appropriate  
heavy-quark limit. They involve  
only three unknown parameters, which may be obtained by fitting to
lattice data. Recent dynamical fermion results from CP-PACS
and UKQCD are extrapolated using these new functional forms. We also
use these functions to probe the limit of applicability of chiral
perturbation theory to the extrapolation of lattice QCD results.
\end{abstract}


\newpage

\section{Introduction}

In the last year there has been tremendous progress in the computation
of baryon masses within lattice QCD. Improved quark
\cite{quark-action} and gluon \cite{gluon-action} actions, together
with increasing computer speed means that one already has results for
$N$, $\Delta$ and vector meson masses for full QCD with two
flavors of dynamical quarks.  Although the results are mainly in the
regime where the pion mass $(m_{\pi })$ is above 500 MeV, there
has been some exploration as low as 300 - 400 MeV on a 3.0 fm lattice
by CP-PACS \cite{CP-PACSlight}.

In spite of these impressive developments it is still necessary to
extrapolate the calculated results to the physical pion mass ($\mu
=140$ MeV) in order to make a comparison with experimental data. In
doing so one necessarily encounters some non-linearity in the quark
mass (or $m_{\pi }^{2}$), including the non-analytic behavior
associated with dynamical chiral symmetry breaking.  Indeed, the recent
CP-PACS study \cite{CP-PACSchi} did report the first behavior of this
kind in baryon systems.

As the computational resources necessary to include three light
flavors with realistic masses will not be available for many years,
it is vital to develop a sound understanding of how to extrapolate to
the physical pion mass. We recently investigated this problem for the
case of the nucleon magnetic moments \cite{Leinweber:1998ej}.

The cloudy bag model (CBM) \cite{Thomas:1984kv} is an extension of the
MIT bag model incorporating chiral symmetry. It therefore generates
the same leading non-analytic (LNA) behavior as chiral perturbation
theory ($\chi$PT).  This model was recently generalized to allow for
variable quark and pion masses in order to explore the likely mass
dependence of the magnetic moment \cite{Leinweber:1998ej}.  This work
led to several important results :

\begin{itemize}
\item a series expansion of $\mu _{p(n)}$ in powers of $m_{\pi }$ is
not a useful approximation for $m_{\pi }$ larger than the physical
mass,
\item on the other hand, the behavior of the model, after adjustments
to fit the lattice data at large $m_{\pi }$ was well determined by
the simple Pad\'e approximant:
\begin{equation}
\label{mag-mom}
\mu _{p(n)}=\frac{\mu _{0}}{1+\frac{\alpha }{\mu _{0}}m_{\pi }+\beta
m^{2}_{\pi }} .
\end{equation}

\item Eq.(\ref{mag-mom}) not only builds in the Dirac moment at
moderately large $m^{2}_{\pi }$ but has the correct LNA behavior of
chiral perturbation theory
\[
\mu =\mu _{0}-\alpha m_{\pi} , 
\]
with $\alpha$ a model independent constant, as $m_{\pi}^2 \rightarrow
0$.
\item fixing $\alpha$ at the value given by chiral perturbation theory
and adjusting $\mu _{0}$ and $\beta$ to fit the lattice data yielded
values of $\mu _{p}$ and $\mu _{n}$ of $2.85\pm 0.22\, \mu _{N}$ and
$-1.96\pm 0.16\, \mu _{N}$, respectively, at the physical pion
mass. These are significantly closer to the experimental values than
the usual linear extrapolations in $m_{q}$.
\end{itemize}
Clearly it is vital to extend the lattice calculations of baryon
magnetic moments to lower values of $m_{\pi }$ than the 600 MeV
used in the study just outlined.  It is also important to include
dynamical quarks. Nevertheless, the apparent success of the
extrapolation procedure suggested by the CBM study gives us strong
encouragement to investigate the same approach for baryon masses.

Accordingly, in this paper we study the variation of the $N$ and
$\Delta$ masses with $m_{\pi }$ (or equivalently $m_{q}$).  Section
\ref{Analyticity-sec} is devoted to considerations of the low-lying
singularities and pion-induced cuts in the complex plane of the
nucleon and $\Delta$ spectral representation.  The analytic properties
of the derived phenomenological form are consistent with both chiral
perturbation theory and the expected behavior at large $m_{q}$. This
phenomenological form is eventually fitted to recent two-flavor, full
QCD measurements made by CP-PACS \cite{CP-PACSlight} and UKQCD
\cite{Allton:1998gi}. However, to gain some insight into the
parameters and behavior of the functional form we examine the $N$ and
$\Delta$ masses as described in the CBM in section
\ref{cbm-calc-sec}. In section \ref{lattice-fit-sec} we apply the
analytic form to the lattice data. Section \ref{discussion-sec} is
reserved for a summary of our findings.

\section{Analyticity}

\label{Analyticity-sec}

\begin{figure}[t]
\centering{\
\epsfig{file=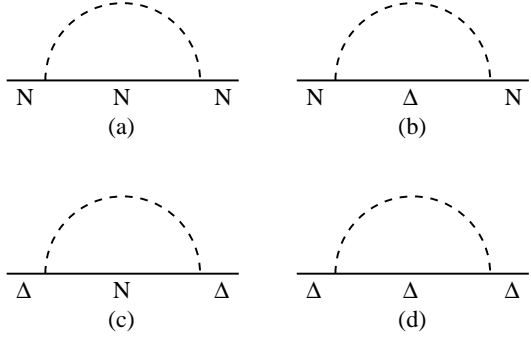,width=7cm}}
\caption{One-loop pion induced self-energy of the nucleon and the
delta.
\label{SE-fig}}
\end{figure}

By now it is well established that chiral symmetry is dynamically
broken in QCD and that the pion is almost a Goldstone boson. As a
result it is strongly coupled to baryons and therefore plays a
significant role in the $N$ and $\Delta$ self energies. In the
limit where the baryons are heavy, the pion induced self-energies of the
$N$ and $\Delta$, to one loop, are given by the processes
shown in Fig.~\ref{SE-fig}. Note that we have restricted the
intermediate baryon states to those most strongly coupled, namely the
$N$ and $\Delta$ states.

The analytic expression for the pion cloud correction to the masses of
the $N$ and $\Delta$ are of the form \cite{Thomas:1999mu}
\begin{equation}
\label{Init-N-eqn}
\delta M_{N}=\sigma _{NN}+\sigma _{N\Delta }\, ,
\end{equation}
where
\begin{eqnarray}
\sigma _{NN} & = & -\frac{3}{16\pi ^{2}f^{2}_{\pi }}g_{A}^{2}\int _{0}^{\infty }dk\frac{k^{4}u^{2}_{NN}(k)}{w^{2}(k)}\, ,\label{SE-NN-integral-eqn} \\
\sigma _{N\Delta } & = & -\frac{3}{16\pi ^{2}f^{2}_{\pi }}\frac{32}{25}g_{A}^{2}\int ^{\infty }_{0}dk\frac{k^{4}u^{2}_{N\Delta }(k)}{w(k)(\Delta M+w(k))}\, ,\label{SE-ND-integral-eqn} 
\end{eqnarray}
and
\begin{equation}
\label{Init-D-eqn}
\delta M_{\Delta }=\sigma _{\Delta \Delta }+\sigma _{\Delta N}\, ,
\end{equation}
where
\begin{eqnarray}
\sigma _{\Delta \Delta } & = & \sigma _{NN}\, ,\\
\sigma _{\Delta N} & = & \frac{3}{16\pi ^{2}f^{2}_{\pi }}\frac{8}{25}g_{A}^{2}\int _{0}^{\infty }dk\frac{k^{4}u^{2}_{N\Delta }(k)}{w(k)(\Delta M-w(k))}\, .\label{SE-DN-integral-eqn} 
\end{eqnarray}
We note that $\Delta M=M_{\Delta }-M_{N}$ , $g_{A}=1.26$ is the axial
charge of the nucleon, $w(k)=\sqrt{k^{2}+m^{2}_{\pi }}$ is the pion
energy and $u_{NN}(k)$, $u_{N\Delta }(k)$, $\ldots$ are the $NN\pi$,
$N\Delta \pi$, $\ldots$ form factors associated with the emission of a
pion of three momentum $k$. We have used SU(6) symmetry to relate
the four coupling constants to the $NN\pi$ coupling, which, in turn,
has been related to $g_A/2f_\pi$ by chiral symmetry.  The form factors
arise naturally in any chiral quark model because of the finite size
of the baryonic source of the pion field -- which suppresses the
emission probability at high virtual pion momentum. As a result, the
self-energy integrals are not divergent.

The leading non-analytic contribution (LNAC) of these self-energy
diagrams is associated with the infrared behavior of the
corresponding integrals -- i.e., the behavior as $k\rightarrow 0$. As
a consequence, the leading non-analytic behavior should not depend on
the details of the high momentum cut-off, or the form factors. In
particular, it should be sufficient for studying the LNAC to evaluate
the self-energy integrals using a simple sharp cut-off, $u(k)=\theta
(\Lambda -k)$.  In Section \ref{cbm-calc-sec} we shall compare the
results with those calculated using a phenomenological, dipole form
factor and show that this is in fact an effective simplification.

Using a $\theta$-function for the form factors, the $NN\pi$
and $\Delta \Delta \pi$ integrals (c.f. Figs. \ref{SE-fig}(a) and
\ref{SE-fig}(d), respectively), which are equal, are easily evaluated
in the heavy baryon approximation used here:
\begin{eqnarray}
\sigma _{NN} = \sigma_{\Delta \Delta} 
 & = & -\frac{3}{16\pi ^{2}f^{2}_{\pi }}g_{A}^{2}\int
 ^{\Lambda }_{0}dk\frac{k^{4}}{w^{2}(k)}\nonumber \\ & = &
 -\frac{3g_{A}^{2}}{16\pi ^{2}f^{2}_{\pi }}\left( m_{\pi }^{3}\arctan
 \left( \frac{\Lambda }{m_{\pi }}\right) +\frac{\Lambda
 ^{3}}{3}-\Lambda m_{\pi }^{2}\right) \, .
\label{NN-Form-eqn}
\end{eqnarray}
The integral corresponding to the process shown in
Fig.\ref{SE-fig}(b), with a $\theta$-function form factor, may be
analytically evaluated.  For $m_\pi > \Delta M$
\begin{eqnarray}
\sigma _{N\Delta } & = & -\frac{g_{A}^{2}}{25\pi ^{2}f^{2}_{\pi}}
 (12(m_{\pi }^{2}-\Delta M^{2})^{3/2}\left\{ \arctan \left(
 \frac{\sqrt{m_{\pi }^{2}+\Lambda ^{2}}+\Delta M+\Lambda
 }{\sqrt{m_{\pi }^{2}-\Delta M^{2}}}\right) \right. \nonumber \\ 
 & &
 -\left. \arctan \left( \frac{\Delta M+m_{\pi }}{\sqrt{m_{\pi }^{2}-\Delta
 M^{2}}}\right) \right\} \nonumber \\ 
 & & +3\Delta
 M(3m_{\pi }^{2}-2\Delta M^{2})\ln \left( \frac{\sqrt{m_{\pi
 }^{2}+\Lambda ^{2}}+\Lambda }{m_{\pi }}\right) \nonumber \\ 
 & &
 -3\sqrt{m_{\pi }^{2}+\Lambda ^{2}}\Delta M\Lambda +6\Delta
 M^{2}\Lambda -6m_{\pi }^{2}\Lambda +2\Lambda ^{3})\, ,
\label{reduced-ND-form}
\end{eqnarray}
while for $m_\pi < \Delta M$ we find
\begin{eqnarray}
\sigma _{N\Delta } & = & -\frac{g_{A}^{2}}{25\pi ^{2}f^{2}_{\pi}}
 (-6(\Delta M^{2}-m_{\pi
 }^{2})^{3/2}\left [\ln \left( \frac{\sqrt{\Delta M^{2}-m_{\pi
 }^{2}}+\sqrt{m_{\pi }^{2}+\Lambda ^{2}}+\Delta M+\Lambda
 }{\sqrt{\Delta M^{2}-m_{\pi }^{2}}-\sqrt{m_{\pi }^{2}+\Lambda
 ^{2}}-\Delta M-\Lambda }\right) \right . \nonumber \\ 
 & & -  \left . \ln \left(
 \frac{\sqrt{\Delta M^{2}-m_{\pi }^{2}}+\Delta M+m_{\pi
 }}{\sqrt{\Delta M^{2}-m_{\pi }^{2}}-\Delta M-m_{\pi }}\right)
 \right ]\nonumber \\ 
 & & + 3\Delta M(3m_{\pi }^{2}-2\Delta M^{2})\ln \left(
 \frac{\sqrt{m_{\pi }^{2}+\Lambda ^{2}}+\Lambda }{m_{\pi }}\right)
 \nonumber \\ 
 & & -  3\sqrt{m_{\pi }^{2}+\Lambda ^{2}}\Delta M\Lambda
 +6\Delta M^{2}\Lambda -6m_{\pi }^{2}\Lambda +2\Lambda ^{3}) \, .
\label{low-ND-form}
\end{eqnarray}
Similar results are easily obtained for the process shown in
Fig.\ref{SE-fig}(c).  For $m_\pi > \Delta M$, the analytic form is
\begin{eqnarray}
\sigma _{\Delta N} & = & \frac{g_{A}^{2}}{100\pi ^{2}f_{\pi}^{2}}
 (-12(m_{\pi }^{2}-\Delta M^{2})^{3/2}\left\{ \arctan \left
 ( \frac{\sqrt{m_{\pi }^{2}+\Lambda ^{2}}-\Delta M+\Lambda
 }{\sqrt{m_{\pi }^{2}-\Delta M^{2}}}\right) \right. \nonumber \\ & &
 \left. +\arctan \left( \frac{\Delta M-m_{\pi }}{\sqrt{m_{\pi
 }^{2}-\Delta M^{2}}}\right) \right\} \nonumber \\ & & +3\Delta
 M(3m_{\pi }^{2}-2\Delta M^{2})\ln \left( \frac{\sqrt{m_{\pi
 }^{2}+\Lambda ^{2}}+\Lambda }{m_{\pi }}\right) \nonumber \\ & &
 -3\sqrt{m_{\pi }^{2}+\Lambda ^{2}}\Delta M\Lambda -6\Delta
 M^{2}\Lambda +6m_{\pi }^{2}\Lambda -2\Lambda ^{3})\, ,
\label{reduced-DN-form}
\end{eqnarray}
while for $m_\pi < \Delta M$
\begin{eqnarray}
\sigma _{\Delta N} & = & \frac{g_{A}^{2}}{100\pi ^{2}f_{\pi}^{2}}
 ( 6(\Delta M^{2}-m_{\pi
 }^{2})^{3/2}\left [\ln \left( \frac{\sqrt{\Delta M^{2}-m_{\pi
 }^{2}}+\sqrt{m_{\pi }^{2}+\Lambda ^{2}}-\Delta M+\Lambda
 }{\sqrt{\Delta M^{2}-m_{\pi }^{2}}-\sqrt{m_{\pi }^{2}+\Lambda
 ^{2}}+\Delta M-\Lambda }\right) \right . \nonumber \\ 
 & & + \left . \ln \left(
 \frac{\sqrt{\Delta M^{2}-m_{\pi }^{2}}+\Delta M-m_{\pi
 }}{\sqrt{\Delta M^{2}-m_{\pi }^{2}}-\Delta M+m_{\pi }}\right)
 \right ]\nonumber \\ & & + 3\Delta M(3m_{\pi }^{2}-2\Delta M^{2})\ln \left(
 \frac{\sqrt{m_{\pi }^{2}+\Lambda ^{2}}+\Lambda }{m_{\pi }}\right)
 \nonumber \\ & & - 3\sqrt{m_{\pi }^{2}+\Lambda ^{2}}\Delta M\Lambda
 -6\Delta M^{2}\Lambda +6m_{\pi }^{2}\Lambda -2\Lambda ^{3}) \, .
\label{low-DN-form}
\end{eqnarray}
The self-energies involving transitions of $N \to \Delta$ or $\Delta
\to N$ are characterized by the branch point at $m_\pi = \Delta M$.

\subsection{Chiral Limit\label{chirlim}}

Chiral perturbation theory is concerned with the behavior of
quantities such as the baryon self-energies as $m_{q}\rightarrow
0$. For the expressions derived above, this corresponds to taking the
limit $m_{\pi }\rightarrow 0$.  The leading non-analytic (LNA) terms
are those which correspond to the lowest order non-analytic functions
of $m_{q}$ -- i.e., odd powers or logarithms of $m_{\pi }$. By
expanding the expressions given above, we find that the LNA
contribution to the nucleon/Delta mass (Eq.(\ref{NN-Form-eqn})) is
given by
\begin{equation}
M_{N(\Delta )}^{LNA}=-\frac{3}{32\pi f^{2}_{\pi }}g_{A}^{2}m^{3}_{\pi
 }\, ,
\label{N:LNA}
\end{equation}
in agreement with a well known result of $\chi$PT \cite{HBchiPT}. A
careful expansion of the $\Delta \pi$ contribution to the nucleon
self energy, Eq.(\ref{reduced-ND-form}), yields the LNA term:
\begin{equation}
\sigma _{N\Delta }(m_\pi,\Lambda ) \sim
\frac{3g_{A}^{2}}{16\pi ^{2}f_{\pi}^{2}}\frac{32}{25}\frac{3}{8\Delta
M}m_{\pi }^{4}\ln (m_{\pi }) 
\label{ND:LNA}
\end{equation}
as $m_\pi \to 0$ which is again as expected from $\chi$PT
\cite{Lebed94}. For the $N\pi$ contribution to the self-energy of the
$\Delta$, the LNA term in the chiral limit of
Eq. (\ref{reduced-DN-form}) yields:
\begin{equation}
\sigma _{\Delta N}(m_\pi,\Lambda ) \sim
 -\frac{3g_{A}^{2}}{16\pi ^{2}f_{\pi
}^{2}}\frac{8}{25}\frac{3}{8\Delta M}m_{\pi }^{4}\ln (m_{\pi })\, .
\label{DN:LNA}
\end{equation}

Of course, our concern with respect to lattice QCD is not so much the
behavior as $m_{\pi }\rightarrow 0$, but the extrapolation from
high pion masses to the physical pion mass. In this context the branch
point at $m_{\pi }^{2}=\Delta M^{2}$ is at least as important as
the LNA near $m_{\pi }=0$. We shall return to this point later. We
note that Banerjee and Milana \cite{Banerjee:1996wz} found the same
non-analytic behavior as $m_{\pi }\rightarrow \Delta M$ that we
find. However they were not concerned with finding a form that could
be used at large pion masses -- i.e. one that is consistent with heavy
quark effective theory.

\subsection{Heavy Quark Limit}

Heavy quark effective theory suggests that as $m_{\pi }\rightarrow
\infty$ the quarks become static and hadron masses become
proportional to the quark mass. This has been rather well explored in
the context of successful nonrelativistic quark models of charmonium
and bottomium\cite{HQET}. In this spirit, corrections are expected to
be of order $1/m_{q}$ where $m_{q}$ is the heavy quark
mass. Thus we would expect the pion induced self energy to vanish as
$1/m_{q}$ as the pion mass increases. The presence of a fixed
cut-off $\Lambda$ acts to suppress the pion induced self energy
for increasing pion masses, as evidenced by the $m_{\pi }^{2}$ in
the denominators of Eqs. (\ref{SE-NN-integral-eqn}),
(\ref{SE-ND-integral-eqn}) and (\ref{SE-DN-integral-eqn}). While some
$m_{\pi }^{2}$ dependence in $\Lambda$ is expected, this is a
second-order effect and does not alter the qualitative features. By
expanding the $\arctan \left( \Lambda / m_\pi \right)$ term in
Eq. (\ref{NN-Form-eqn}) for small $\Lambda/m_\pi$, we find
\begin{equation}
\label{NN-heavy-lim}
\sigma_{NN}=-\frac{3g_{A}^{2}}{16\pi ^{2}f^{2}_{\pi }}\frac{\Lambda
^{5}}{5m^{2}_{\pi }}+{\cal O}\left( \frac{\Lambda ^{7}}{m^{4}_{\pi
}}\right) \, ,
\end{equation}
which vanishes for $m_{\pi }\rightarrow \infty$. Indeed, in the
large $m_{\pi }$ (heavy quark) limit, both
Eqs. (\ref{reduced-ND-form}) and (\ref{reduced-DN-form}) tend to zero
as $1/m_{\pi }^{2}$.

\subsection{Analytic form}

We now have the chiral and heavy quark limits for each of the four
integrals in Fig.\ref{SE-fig}. These expressions, which contain a
single parameter, $\Lambda$, are correct in the chiral limit - i.e.,
they reproduce the first two non-analytic terms of $\chi$PT. They also
have the correct behavior in the limit of large pion mass, namely
they vanish like $1/m_{\pi }^{2}$.  The latter feature would be
destroyed if we were to retain only the LNA pieces of the
self-energies as they would diverge at large $m_{\pi }$ faster than
$m_{q}$. Rather than simplifying our expressions to just the LNA
terms, we therefore retain the complete expressions, as they contain
important physics that would be lost by making a simplification.

We note that keeping the entire form is not in contradiction with
$\chi$PT, as we have already shown that the leading non-analytic
structure of $\chi$PT is contained in this form. However as one
proceeds to larger quark (pion) masses, differences between the full
forms and the expressions in the chiral limit will become
apparent. For example, the branch point at $m^{2}_{\pi }=\Delta
M^{2}$, which is an essential non-analytic component of the $m_{\pi
}$-dependence of the self-energy and which should dominate in the
region $m_{\pi }\sim \Delta M$, is also satisfactorily incorporated in
Eqs.(\ref{reduced-ND-form}) and (\ref{reduced-DN-form}).  Yet the LNA
chiral terms given in section \ref{chirlim} know nothing of this
branch point and are clearly inappropriate in the region near and
beyond $m^{2}_{\pi }=\Delta M^{2}$.

As a result of these considerations, we propose to use the analytic
expressions for the self-energy integrals corresponding to a sharp
cut-off in order to incorporate the correct LNA structure in a simple
three-parameter description of the $m_{\pi }$-dependence of the $N$
and $\Delta$ masses. In the heavy quark limit hadron masses become
proportional to the quark mass.  Moreover, as we shall see in the next
section, the MIT bag model leads to a linear dependence of the mass of a
baryon on the current quark mass far below the scale at which one would
expect the heavy quark limit to apply. This is a simple consequence of
relativistic quantum mechanics for a scalar confining field. On the
other hand, lattice calculations indicate that the
scale at which the pion mass exhibits a linear dependence on 
$m_q$ is much larger than
that for baryons.\footnote{One doesn't expect such linear behavior to
appear for quark masses lighter than the charm quark mass where the
pseudoscalar mass is 3.0 GeV.  Even at this scale the quarks are still
somewhat relativistic.} In fact, over the range of masses of interest to
us, explicit lattice calculations show that 
$m_\pi^2$ is proportional to $m_q$. 
Hence we can simulate a linear dependence of
the baryon masses on the quark mass, $m_q$, in this region,
by adding a term involving $m^{2}_{\pi }$.
The functional form for the mass of the nucleon suggested by this
analysis is then:
\begin{equation}
\label{N-form-eqn}
M_{N}=\alpha _{N}+\beta _{N}m_{\pi }^{2}+\sigma _{NN}(m_{\pi },\Lambda
)+\sigma _{N\Delta }(m_{\pi },\Lambda )\, ,
\end{equation}
while that for the $\Delta$ is:
\begin{equation}
\label{D-form-eqn}
M_{\Delta }=\alpha _{\Delta }+\beta _{\Delta }m_{\pi }^{2}+\sigma
_{\Delta \Delta }(m_{\pi },\Lambda )+\sigma _{\Delta N}(m_{\pi
},\Lambda )\, .
\end{equation}
The mass in the chiral limit is given by
\begin{equation}
M_{N}^{(0)}=\alpha _{N}+\sigma _{NN}(0,\Lambda )+\sigma _{N\Delta
}(0,\Lambda )\, ,
\end{equation}
where the meson cloud effects are explicitly contained in $\sigma
_{NN}(0,\Lambda )+\sigma _{N\Delta }(0,\Lambda )$. The mass of the
$\Delta$ in the chiral limit is calculated in an analogous way. We
know that (\ref{N-form-eqn}) and (\ref{D-form-eqn}) have the correct
behavior in the chiral limit.  Individually, they also have the
correct heavy quark behavior.\footnote{With regard to the difference,
$M_\Delta - M_N$, HQET suggests that this difference should vanish as
$m_\pi \to \infty$.  This is only guaranteed by
Eqs. (\protect\ref{N-form-eqn}) and (\protect\ref{D-form-eqn})
(through Eq. (\protect\ref{NN-heavy-lim})) if the entire mass
difference arises from the pion self-energy.  While one could enforce
this condition through the introduction of additional parameters and a
more complicated analytic structure for the higher-order terms of
Eqs. (\protect\ref{N-form-eqn}) and (\protect\ref{D-form-eqn}), we
prefer to focus on the regime of $m_\pi^2$ from 1 GeV${}^2$ to the
chiral limit.  As we shall see, Eqs. (\protect\ref{N-form-eqn}) and
(\protect\ref{D-form-eqn}) are quite adequate for this purpose.}
Between the chiral and heavy-quark limits there are no general
guidelines, so in the next section we shall compare our functional
form to the Cloudy Bag Model, a successful phenomenological approach
incorporating chiral symmetry and the correct heavy quark limit.

\section{Baryon Masses within the CBM}

\label{cbm-calc-sec}

As a guide to the quark mass dependence of the $N$ and $\Delta$ masses
we consider the Cloudy Bag Model (CBM)
\cite{Thomas:1984kv,Theberge:1980ye}.  This is a minimal extension of
the MIT bag model such that chiral symmetry is restored, which has
proven quite successful in a number of phenomenological studies of
baryon properties and meson-baryon scattering
\cite{Thomas:1984kv,Lu:1997sd,Lu:1998se,Veit:1985sf}.
Within the CBM, a baryon is viewed as a superposition of a bare quark
core and bag plus meson states. The linearized CBM Lagrangian with
pseudovector pion-quark coupling (to order $1/f_{\pi }$) 
is \cite{VCBM}:
\begin{eqnarray}
\cal {L} & = & \left[ \bar{q}(i\gamma ^{\mu }\partial _{\mu
 }-m_{q})q-B\right] \theta _{V}-\frac{1}{2}\bar{q}q\delta
 _{S}\nonumber \\ & & +\frac{1}{2}(\partial _{\mu }\bfgreek {\pi
 })^{2}-\frac{1}{2}m^{2}_{\pi }\bfgreek {\pi }^{2} + 
 \frac{\theta _{V}}{2f_{\pi
 }}\bar{q} \gamma^\mu \gamma _{5} \bfgreek {\tau } q 
 \cdot \partial_\mu \bfgreek {\pi }
 \, ,
\end{eqnarray}
where $B$ is the bag constant, $f_{\pi }$ is the $\pi$
decay constant, $\theta _{V}$ is a step function (unity inside the
bag volume and vanishing outside) and $\delta _{S}$ is a surface
delta function. In a lowest order perturbative treatment of the pion
field, the quark wave function is not effected by the pion field and
is simply given by the MIT bag solution
\cite{Chodos:1974je,Chodos:1974pn,DeGrand:1975cf}. 

In principle the $\pi NN$ form factor can be directly calculated
within the model. It dies off at large momentum transfer because of
the finite size of the baryon source. Rather than using this
calculated form factor, which is model dependent, we have chosen to
use a common phenomenological form, namely a simple dipole:
\begin{equation}
u(k)=\frac{(\Lambda _{D}^{2}-\mu ^{2})^{2}}{(\Lambda
_{D}^{2}+k^{2})^{2}}\, ,
\end{equation}
where $k$ is the magnitude of the loop (3-)momentum, $\mu$ is
the physical pion mass (139.6 MeV), and $\Lambda _{D}$ is a
regulation parameter.

In the standard CBM treatment, where the pion is treated as an
elementary field, the current quark mass, $m_{q}$, is not directly
linked to $m_{\pi }$.  Most observables are not sensitive to this
parameter, as long as it is in the range of typical current quark
masses. For our present purpose it is vital to relate the $m_{q}$
inside the bag with $m_{\pi }$. Current lattice simulations indicate
that $m_{\pi }^{2}$ is approximately proportional to $m_{q}$ over a
wide range of quark masses \cite{CP-PACSlight}. Hence, in order to
model the lattice results, we scale the mass of the quark confined in
the bag as $m_{q}=\left( m_{\pi }/\mu \right) ^{2}m_{q}^{(0)}$, with
$m_{q}^{(0)}$ being the current quark mass corresponding to the
physical pion mass $\mu$. $m_{q}^{(0)}$ is treated as an input
parameter to be tuned to the lattice results, but in our magnetic
moment study it turned out to lie in the range 6 to 7 MeV, which is
very reasonable.

\begin{figure}[t]
\centering{\
\rotate{\epsfig{file=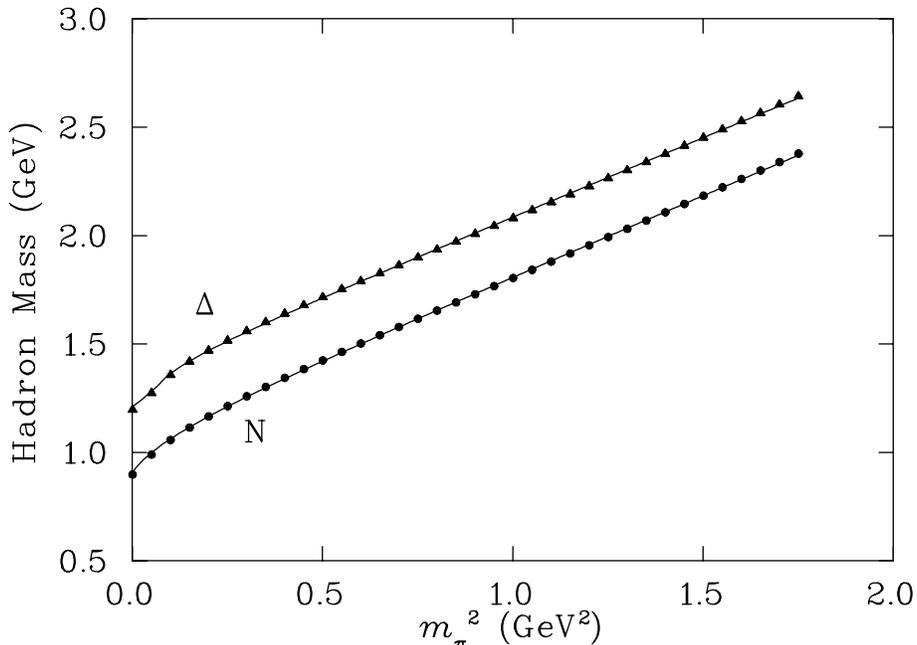,height=12cm}}
\caption{The pion mass dependence of the $N$ and
$\Delta$ baryons generated in the CBM using a
dipole form factor with $\Lambda _{D}=1$ GeV. Fits
of (\ref{N-form-eqn}) and (\ref{D-form-eqn}) to the CBM results are
illustrated by the curves.
\label{CBM-masses-fig}}}
\end{figure}

The parameters of the CBM are obtained as follows. The bag constant
$B$ and the phenomenological parameter $z_{0}$ are fixed by the
physical nucleon mass and the stability condition,\footnote{Note that
while $z_0, B$ and the $\pi NN$ form factor may 
all depend on $m_q$, this dependence is expected to be a smaller
effect and we ignore such variations in order to avoid an excess of
parameters.}  
$dM_{N}/dR=0$, for a given choice of $R_{0}$ and
$m_{q}^{(0)}$. For each subsequent value of the pion mass or the quark
mass considered, $\omega _{0}$ and $R$ are determined simultaneously
from the linear boundary condition
\cite{Chodos:1974je,Chodos:1974pn,DeGrand:1975cf} and the stability
condition.  In this work we have calculated the mass of the $N$ and
$\Delta$ baryons as a function of squared pion mass (as illustrated in
Fig. \ref{CBM-masses-fig}).  The $\Delta$ calculation is similar to
that for the $N$, however the value of $B$ is fixed to be the same as
that used for the nucleon, while $z_{0}$ is adjusted to fit the
observed mass difference, taking into account the pionic contribution
to this quantity, at the physical value $m_{\pi} = \mu$ ($m_{q} =
m_{q}^{(0)}$).

As expected on quite general grounds (and discussed in Section
\ref{Analyticity-sec}), as the pion mass increases the mass of the
baryon does indeed become linear in $m_{\pi }^{2}$. In addition,
from the curvature at low pion mass, we see that the non-analytic
structure is important in the region $m_{\pi }$ below 400 MeV.

We now fit our functional forms for the baryon masses,
Eqs. (\ref{N-form-eqn}) and (\ref{D-form-eqn}), to the CBM data.  We
note that the CBM data is generated using a phenomenologically
motivated, dipole form factor, whereas the functional form used in the
fit involves a $\theta$ cut-off.  In order to simulate the fitting
procedure for lattice data, our fit to the CBM results involves only
pion masses above the physical branch point at $M_{\Delta
}=M_{N}$, followed by an extrapolation to lower pion mass.

It can be seen from Fig. \ref{CBM-masses-fig} that our extrapolation
to the physical pion mass is in good agreement with the CBM
calculations: at the physical pion mass the extrapolated $N$ mass
is within 0.8\% of the experimental value to which the CBM was fitted,
while the $\Delta$ is within 0.3\% of the experimental value. We
present the parameters of our fit in Table \ref{CBM-params-table}.
The value for the sharp cut-off ($\Lambda$) is 0.44(2) GeV,
compared to $\Lambda _{D}=$1 GeV for the dipole form factor.

It was noted in Section \ref{Analyticity-sec} that the constant
$\alpha$ in our functional form is not the mass of the baryon in the
chiral limit, but rather this is given by $M_{N}^{(0)}=\alpha
_{N}+\sigma _{NN}(0,\Lambda )+\sigma _{N\Delta }(0,\Lambda )$ -- with
an analogous expression for the $\Delta$. We find that the
extrapolated $N$ and $\Delta$ masses in the chiral (SU(2)-flavor)
limit are $(M_{N}^{(0)},M^{(0)}_{\Delta }) = (905,1210)$ MeV, compared
with the CBM values (898,1197) MeV.

\begin{figure}[t]
\centering{\
\rotate{\epsfig{file=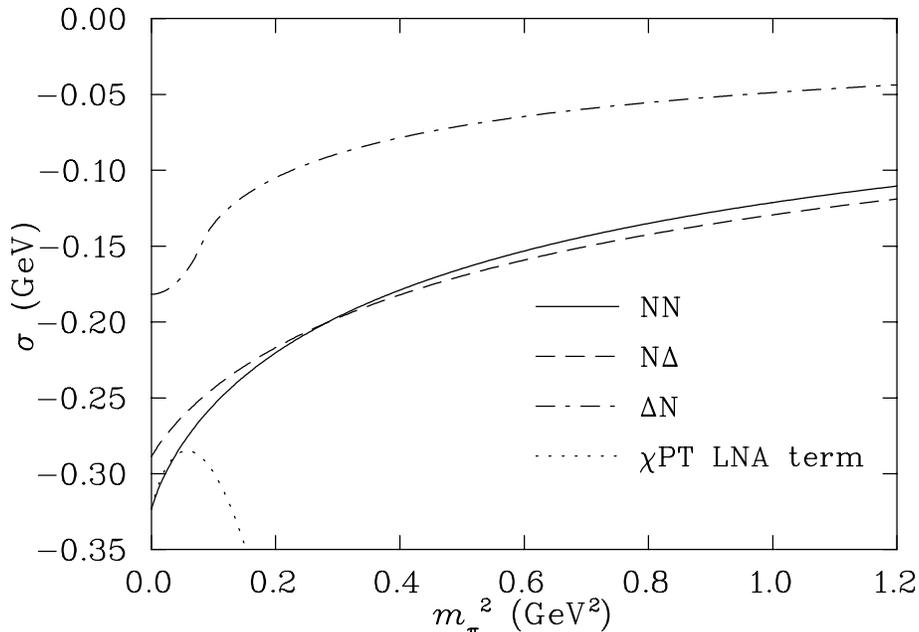,height=12cm}}
\caption{Pion induced self-energy corrections for a 1 GeV dipole form
factor.  The LNA term of $\chi$PT tracks the $NN\pi$ contribution up
to $m_\pi \sim 0.2$ GeV, beyond which the internal structure of the
nucleon becomes important.
\label{SE-Dipole-fig}}}
\end{figure}

\begin{table}[b]
{\centering \begin{tabular}{cccccc}
Baryon&
$\alpha $&
$\beta $&
$\Lambda $&
$M_{B}$ &Error\\
&(GeV) &(GeV${}^{-1}$) &(GeV) &(GeV) \\
\hline 
$N$&
1.09&
0.739&
0.455&
0.948 &0.8\%\\
$\Delta $&
1.37&
0.725&
0.419&
1.236 &0.3\%
\end{tabular}\par}
\caption{Parameters for fitting (\protect\ref{N-form-eqn}) and
(\protect\ref{D-form-eqn}) to the CBM data.  Here we have taken
$R_{0}=1.0$~fm and $m_{q}^{(0)}=6.0$ MeV.  The Error
column denotes the relative difference from the experimental values
which were used as a constraint in generating the CBM data.
\label{CBM-params-table}}
\end{table}

The mass dependence of the pion induced  
self-energies, $\sigma_{i j}$, for the 
1 GeV dipole form factor, is displayed in Fig. \ref{SE-Dipole-fig}.  The
choice of a 1 GeV dipole corresponds to the observed axial form factor
of the nucleon \cite{Axial_data}, which is probably our best
phenomenological guide to the pion-nucleon form factor \cite{guide}.
We note that $\sigma _{NN}$ tends to zero smoothly as $m_\pi$ grows and
it is only below $m_\pi^2 \sim 0.3$ GeV$^2$ that there is any rapid
variation. That this behaviour cannot be well described by a polynomial
expansion is illustrated by the dotted curve in Fig.
\ref{SE-Dipole-fig}. There we expanded $\sigma _{NN}$ about $m_\pi = 0$
as a simple polynomial, $\alpha + \beta m_\pi^2 + \gamma m_\pi^3$, with
$\gamma$ fixed at the value required by chiral symmetry. Clearly the
expansion fails badly for $m_\pi$ beyond 300-400 MeV.

The behavior of the $N\pi$ contribution to the self-energy of the
$\Delta$ is especially interesting. In particular, the effect of the
branch point at $m_{\pi }=\Delta M$ is seen in the curvature at
$m^{2}_{\pi }\sim 0.1\ {\rm GeV}^2$.  For comparison, we note that
while there is also a branch point in the nucleon self-energy at the
same point -- see Eq. (\ref{reduced-ND-form}) -- the coefficient of
$(m_{\pi }^{2}-\Delta M^{2})^{\frac{3}{2}}$ vanishes at this point.
As a consequence there is little or no curvature visible in the latter 
quantity at the same point.  The correct description of this curvature
is clearly very important if one wishes to obtain the $\Delta N$ mass
difference at the physical pion mass. The fact that, as shown in
Fig. \ref{CBM-masses-fig}, our simple three parameter phenomenological
fitting function can reproduce $N$ and $\Delta$ masses within the CBM,
including this curvature, suggests that this should also provide a
reliable form for extrapolating lattice data into the region of small
pion mass.
\begin{figure}[t]
\centering{\
\rotate{\epsfig{file=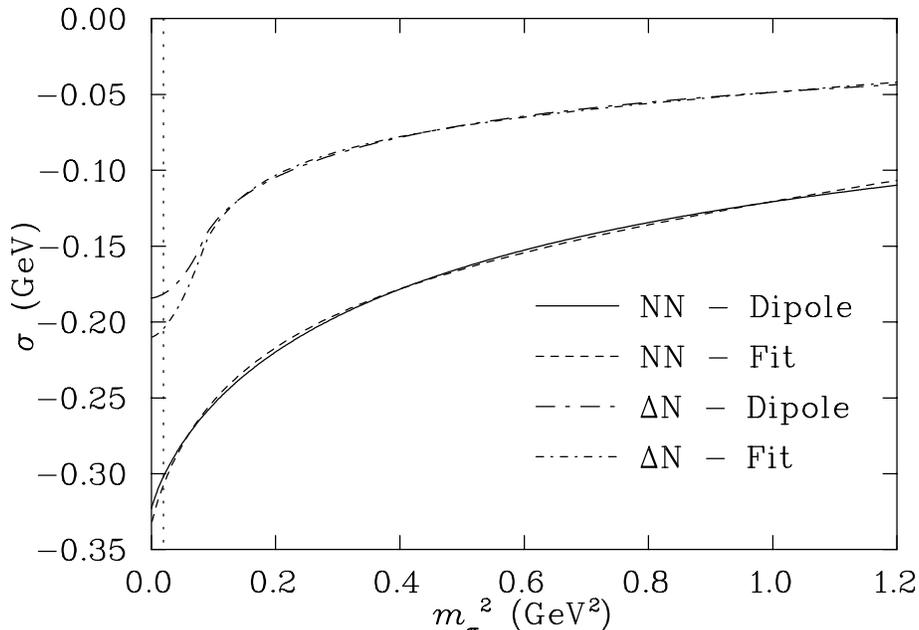,height=12cm}}
\caption{Comparison between the nucleon and $\Delta$ self-energies, 
$\sigma _{NN}$ and $\sigma _{\Delta N}$, calculated using a dipole form
factor (solid and long-dash dot curves, respectively) and fits using
the form $\alpha + \beta m_\pi^2 + \sigma _{ij}(m_\pi, \Lambda)$, based
on a sharp cut-off in the momentum of the virtual pion (dash and short-dash
dotted curves respectively).
\label{SE_Fit-fig}}}
\end{figure}

{}Fig. \ref{SE_Fit-fig} illustrates the degree of residual model dependence in our use
of (17) and (18). There 
the variation of the nucleon self-energy, $\sigma _{NN}$, calculated with 
a 1 GeV dipole form factor (solid curve) is fit using the form $\alpha + \beta
m_\pi^2 + \sigma _{NN}(m_\pi, \Lambda)$ (dash curve, with $\alpha = -0.12$
GeV, $ \beta = 0.39$ GeV$^{-1}$ and $\Lambda = 0.57$ GeV). Note that the
deviations are at the level of a few MeV. For the $\Delta$ the
self-energy, $\sigma _{\Delta N}$, is again calculated using a 1 GeV dipole
form factor and fit with our standard fitting function,  
$\alpha + \beta m_\pi^2 + \sigma _{\Delta N}(m_\pi, \Lambda)$. The
quality of the fit (with $\alpha = -0.062$GeV, $ \beta = 0.024$
GeV$^{-1}$ and $\Lambda = 0.53$ GeV) is not as good as for the 
nucleon case. Nevertheless, the difference between the two curves at the
physical pion mass (vertical dotted line) is only about 20 MeV. At the
present stage of lattice calculations this seems to be an acceptable
level of form factor dependence for such a subtle extrapolation.

\section{Lattice Data Analysis}

\label{lattice-fit-sec}

We consider two independent lattice simulations of the $N$ and
$\Delta$ masses, both of which use improved actions to study baryon
masses in full QCD with two light flavors.  The CP-PACS
\cite{CP-PACSlight} lattice data is generated on a plaquette plus
rectangle gauge action with improvement coefficients based on an
approximate block-spin renormalization group analysis.  The ${\cal
O}(a)$-improved Sheikholeslami and Wohlert clover action is used with
a mean-field improved estimate of the clover coefficient $c_{SW}$=1.64
-- 1.69.  This estimate is likely to lie low relative to a
nonperturbative determination \cite{Jansen:1998mx} and may leave
residual ${\cal O}(a)$ errors.

Ideally one would like to work with lattice data in which the
infinite-volume continuum limit is taken prior to the chiral limit.
Until such data is available, we select results from their $12^3
\times 32$ and $16^3 \times 32$ simulations at $\beta=1.9$.  Lattice
spacings range from 0.25 fm to 0.19 fm and provide physical volumes
2.7 fm to 3.5 fm on a side.  While the volumes are large enough to
avoid significant finite volume errors, the coarse lattice spacings
necessitate the use of improved actions.  Systematic uncertainties the
order of 10\% are not unexpected.

The UKQCD \cite{Allton:1998gi} group uses a standard plaquette action
with the ${\cal O}(a)$-improved Sheikholeslami and Wohlert action.  At
a $\beta$ of 5.2, UKQCD uses $c_{SW}=1.76$, which is lower than the
current non-perturbative value \cite{Jansen:1998mx} of 2.017, again
leaving some residual ${\cal O}(a)$ errors.  Lattice spacings are
necessarily smaller, ranging from 0.13 to 0.21 fm.  We select their
$12^3 \times 24$ data set providing better statistical errors than
their largest volume simulation.  Physical volumes are 1.6 to 2.6 fm
on a side suggesting that finite volume errors may be an issue on the
smallest physical volume where the dynamical quark mass is lightest.

In full QCD, the renormalized lattice spacing is a function of both
the bare coupling and the bare quark mass. In order to determine the
lattice spacing, the UKQCD collaboration calculates the force between 
two static quarks at a distance $r_0$ \cite{sommer94}, while 
CP-PACS considers the string tension directly.
While the two approaches yield
similar results in the quenched approximation, string breaking in full
QCD may introduce some systematic error in the extraction of the
string tension at large distances.  In fact we find that the two data sets
are consistent, provided one allows the parameters introducing the
physical scale to float within systematic errors of 10\%.  A thorough
investigation of these systematic errors lies outside the scope of
this investigation.  Instead we simply rescale the UKQCD and CP-PACS
data sets in combining them into a single, consistent data set.

\begin{figure}[p]
\centering{\
\rotate{\epsfig{file=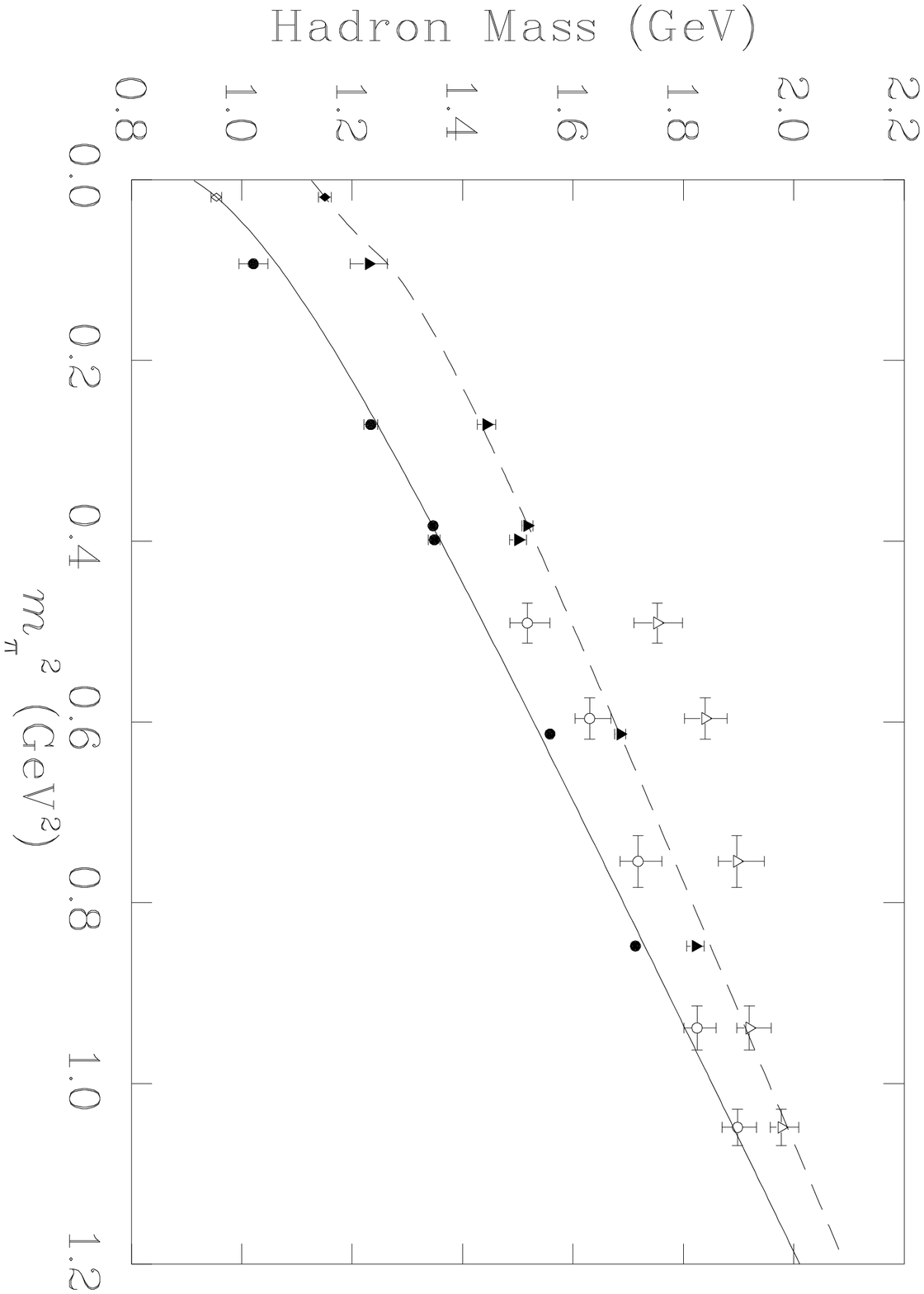,height=12cm}}
\caption{Baryon masses calculated by UKQCD (open points) and CP-PACS
(solid points), as a function of $m_\pi^2$.  The solid (dashed) curve
illustrates a fit to the combined data sets for $N$ ($\Delta$).  The
left-most data points are our extrapolated values of the baryon masses
at the physical pion mass.
\label{Lat+0-0-fig}}}
\end{figure}

\begin{figure}[p]
\centering{\
\rotate{\epsfig{file=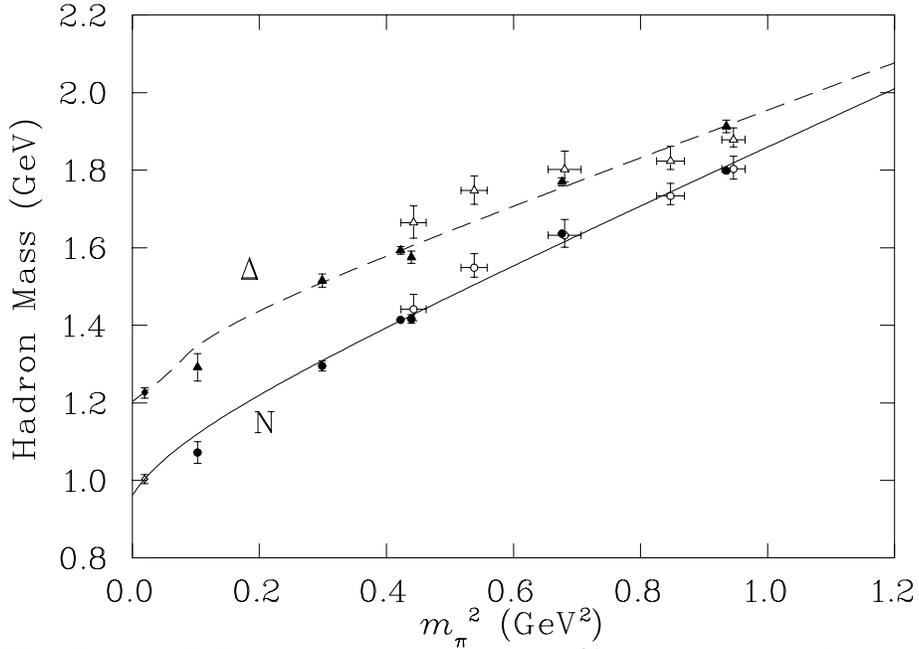,height=12cm}}
\caption{UKQCD and CP-PACS baryon masses with 5\% adjustments in the
scale parameters to improve the agreement between the two data sets.
(The key is as described in Fig. \ref{Lat+0-0-fig}.)
\label{Lat+5-5-fig}}}
\end{figure}

We begin by considering the functional form suggested in Section
\ref{Analyticity-sec} with the cut-off $\Lambda$ fixed to the
value determined by fitting the CBM calculations.  The resulting fits
to the baryon masses are shown in Fig.\ \ref{Lat+0-0-fig} for the
unshifted lattice data and Fig. \ref{Lat+5-5-fig} where each data set
is adjusted by 5\% to provide consistency.  The extrapolations
are indicated by the solid (dashed) curve for $N$ ($\Delta$).  The
resulting fit parameters and masses\footnote{The errors bars for the
extrapolated baryon masses at the physical pion mass displayed in the
figures are naive estimates only.  We are unable to perform a complete
analysis without the lattice results on a configuration by
configuration basis.} are listed in Table \ref{Lat-params-tab}.

\begin{table}[t]
{\centering \begin{tabular}{cccccccccc}
\multicolumn{2}{c}{Scaling }&
\multicolumn{4}{c}{$N$ }&
\multicolumn{4}{c}{$\Delta$ }\\
{\tiny CP-PACS}{\footnotesize }&
{\tiny UKQCD}&
$\alpha $&
$\beta $&
$M_{N}$&
$M_{N}^{(0)}$&
$\alpha $&
$\beta $&
$M_{\Delta }$&
$M_{\Delta }^{(0)}$\\
& &(GeV) &(GeV${}^{-1}$) &(GeV) &(GeV) &(GeV) &(GeV${}^{-1}$) &(GeV) &(GeV) \\
\hline 
0\%&
0\%&
1.10&
0.778&
0.954&
0.910&
1.29&
0.680&
1.150&
1.125\\
 +5\%&
-5\%&
1.15 &
0.736 &
1.003&
0.961&
1.36&
0.602&
1.227&
1.203\\
0\%&
-10\%&
1.10&
0.767&
0.957&
0.914&
1.31&
0.624&
1.169&
1.145\\
+10\%&
0\%&
1.20&
0.707&
1.050&
1.008&
1.42&
0.581&
1.285&
1.262
\end{tabular}\par}
\caption{Parameters for fits of (\protect\ref{N-form-eqn}) and
(\protect\ref{D-form-eqn}) to lattice data.  Here we fix $\Lambda$
($\Lambda _{N}=0.455$ and $\Lambda _{\Delta
}=0.419$) and vary $\alpha$ and
$\beta $. The mass of the baryon at the physical pion
mass is $M_{N}$ ($M_{\Delta }$) and
the mass in the chiral limit is
$M_{N}^{(0)}$ ($M^{(0)}_{\Delta }$).
The scaling columns represent adjustments to the scale parameters
providing physical dimensions to the lattice data.
\label{Lat-params-tab}}
\end{table}

\begin{figure}[p]
\centering{\
\rotate{\epsfig{file=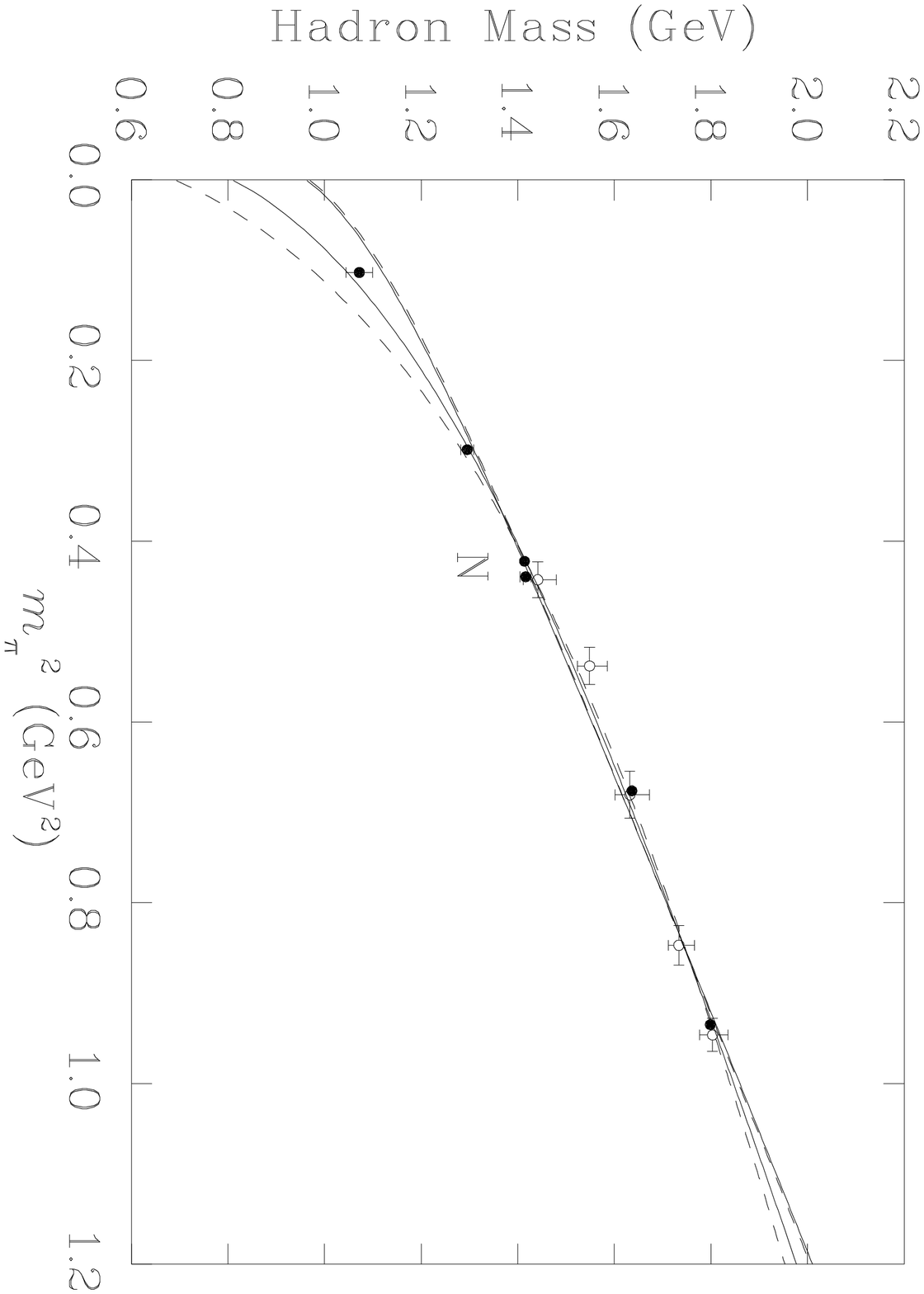,height=12cm}}
\caption{UKQCD and CP-PACS nucleon masses with scale parameters
adjusted by 5\%. The data is as described in
Fig. \ref{Lat+0-0-fig}. The dashed lines represent fits without the
point at 0.1 GeV$^{2}$.  The solid lines include this point.  The top
pair of lines are fits with $\Lambda$ fixed at 0.455 GeV, a value
preferred on the basis of our CBM analysis.  The bottom pair have
$\Lambda$ as a fit parameter.\label{Lat-Vary-Points-N-fig}}}
\end{figure}

\begin{figure}[p]
\centering{\
\rotate{\epsfig{file=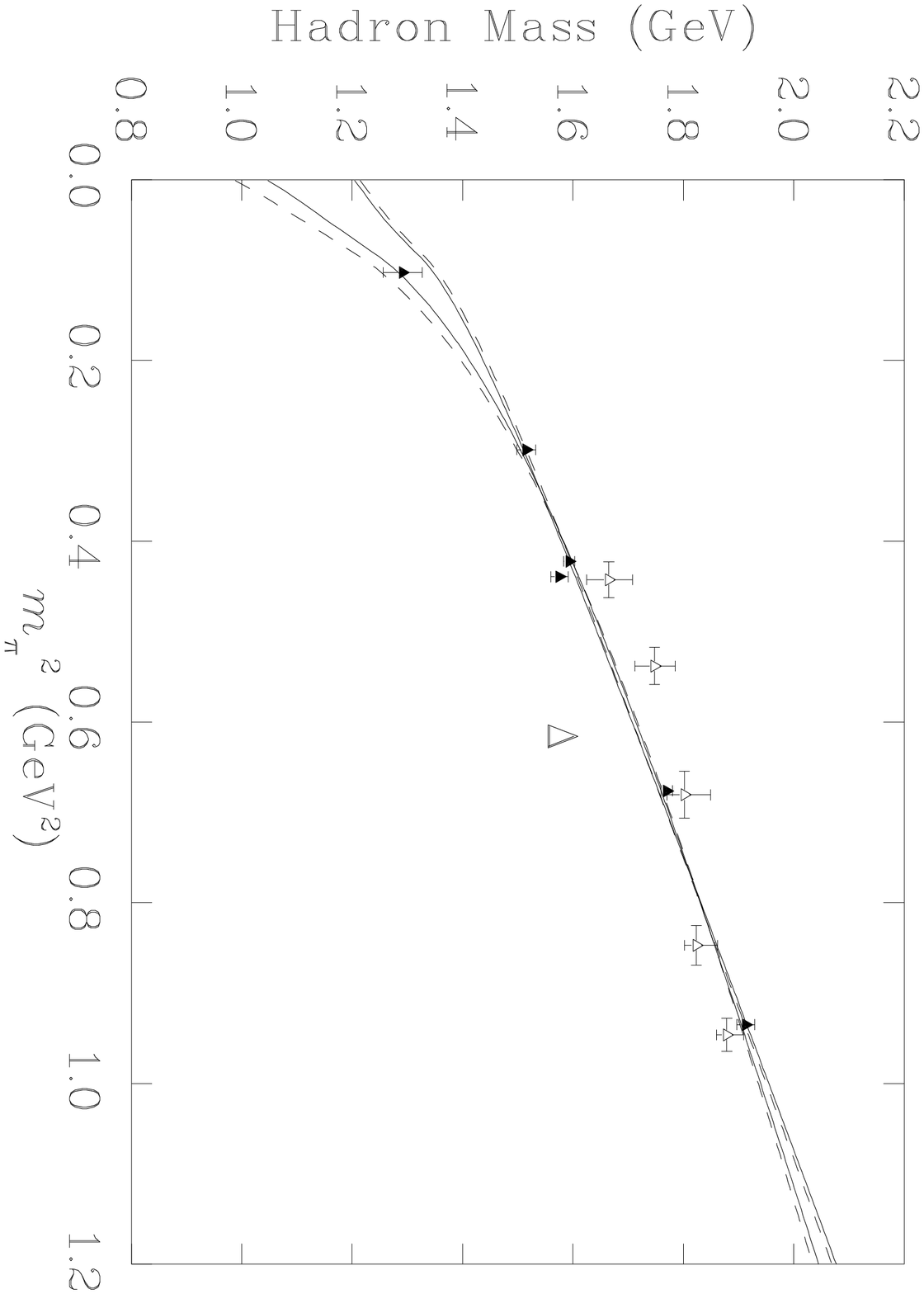,height=12cm}}
\caption{UKQCD and CP-PACS $\Delta$-baryon masses with scale
parameters adjusted by 5\%. The data is as described in
Fig. \ref{Lat+0-0-fig}. The dashed lines represent fits without the
point at 0.1 GeV$^{2}$.  The solid lines include this point.  The top
pair of lines are fits with $\Lambda$ fixed at 0.419 GeV, a value
preferred on the basis of our CBM analysis.  The bottom pair have
$\Lambda$ as a fit parameter.\label{Lat-Vary-Points-D-fig}}}
\end{figure}

In examining fits in which the cutoff is allowed to vary as a fit
parameter, we found it instructive to also study the dependence of the
fit on the number of points included.  This dependence is shown for
the $N$ in Fig. \ref{Lat-Vary-Points-N-fig} and for the $\Delta$
in Fig. \ref{Lat-Vary-Points-D-fig}.
In particular, we compare fits including the lowest lattice point (at
around 0.1 GeV$^{2}$) and then excluding it. When we fix the value of
$\Lambda$ the fits are stable and insensitive to the lowest
point. They tend to lie slightly above the lowest data point. However,
given the caution expressed by the CP-PACS collaboration for the
lowest point, we view these fits as reasonably successful.  In
contrast, when the value of $\Lambda$ is treated as a fitting parameter,
it is sensitive to the inclusion of the lowest point.  Hence, to perform
model independent fits, it is essential to have lattice simulations at
light quark masses approaching $m^{2}_{\pi }\sim 0.1$ GeV$^{2}$.  An
analysis of the current data suggests $\Lambda =0.661$ GeV and
provides a nucleon mass 130 MeV lower than the CBM-constrained fit.
Tables \ref{Lat-Vary-Points-N-table} and \ref{Lat-Vary-Points-D-table}
summarize parameters and physical baryon masses for $N$ and $\Delta$
respectively.

\begin{table}[b]
{\centering \begin{tabular}{ccccc}
Fit &
$\alpha $&
$\beta $&
$\Lambda $&
$M_{N}$\\
&(GeV) &(GeV${}^{-1}$) &(GeV) &(GeV) \\
\hline 
(a)&
1.76&
0.386&
0.789&
0.763\\
(b)&
1.15&
0.727&
0.455&
1.010\\
(c)&
1.42&
0.564&
0.661&
0.870\\
(d)&
1.15&
0.736&
0.455&
1.003\\
\end{tabular}\par}
\caption{Parameters for the fits shown in Fig.\
\ref{Lat-Vary-Points-N-fig}.  Parameter sets (a) and (b) are obtained
by excluding the lowest data point from the fit, while (c) and (d)
include it.  Parameter sets (a) and (c) are fits with 3 parameters,
sets (b) and (d) are fits with $\Lambda$ fixed to
the phenomenologically preferred value.
\label{Lat-Vary-Points-N-table}}
\end{table}

\begin{figure}[p]
\centering{\
\rotate{\epsfig{file=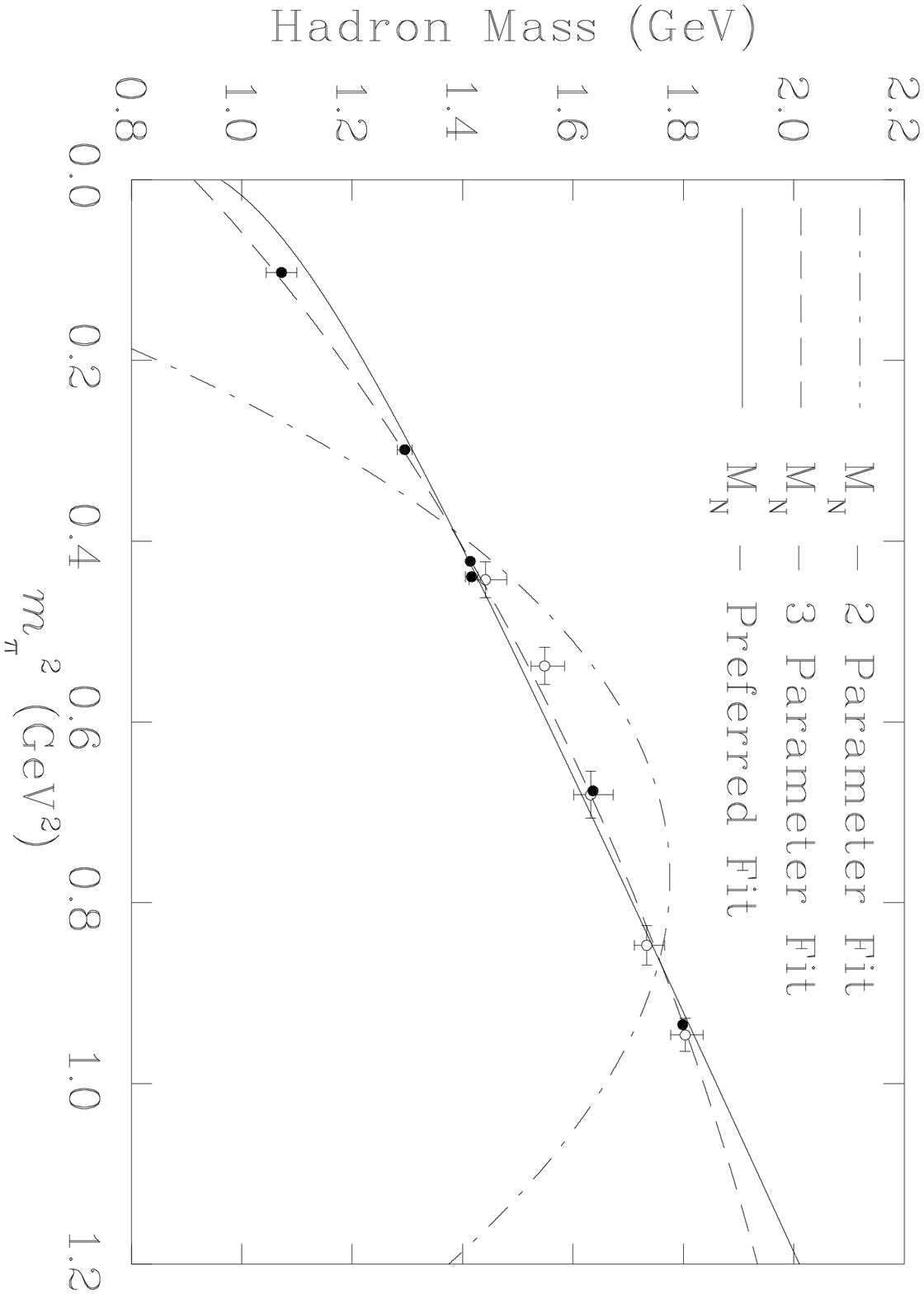,height=12cm}}
\caption{A comparison between phenomenological fitting functions for
the mass of the nucleon.  The two parameter fit corresponds to using
Eq. \ref{Naive-eqn} with $\gamma$ set equal to
the value known from $\chi$PT.  The three
parameter fit corresponds to letting $\gamma $
vary as an unconstrained fit parameter. The solid line is the fit for
the functional form of (\ref{N-form-eqn}), fit (d) of Table
\ref{Lat-Vary-Points-N-table}.
\label{3fits-N-fig}}}
\end{figure}

\begin{figure}[p]
\centering{\
\rotate{\epsfig{file=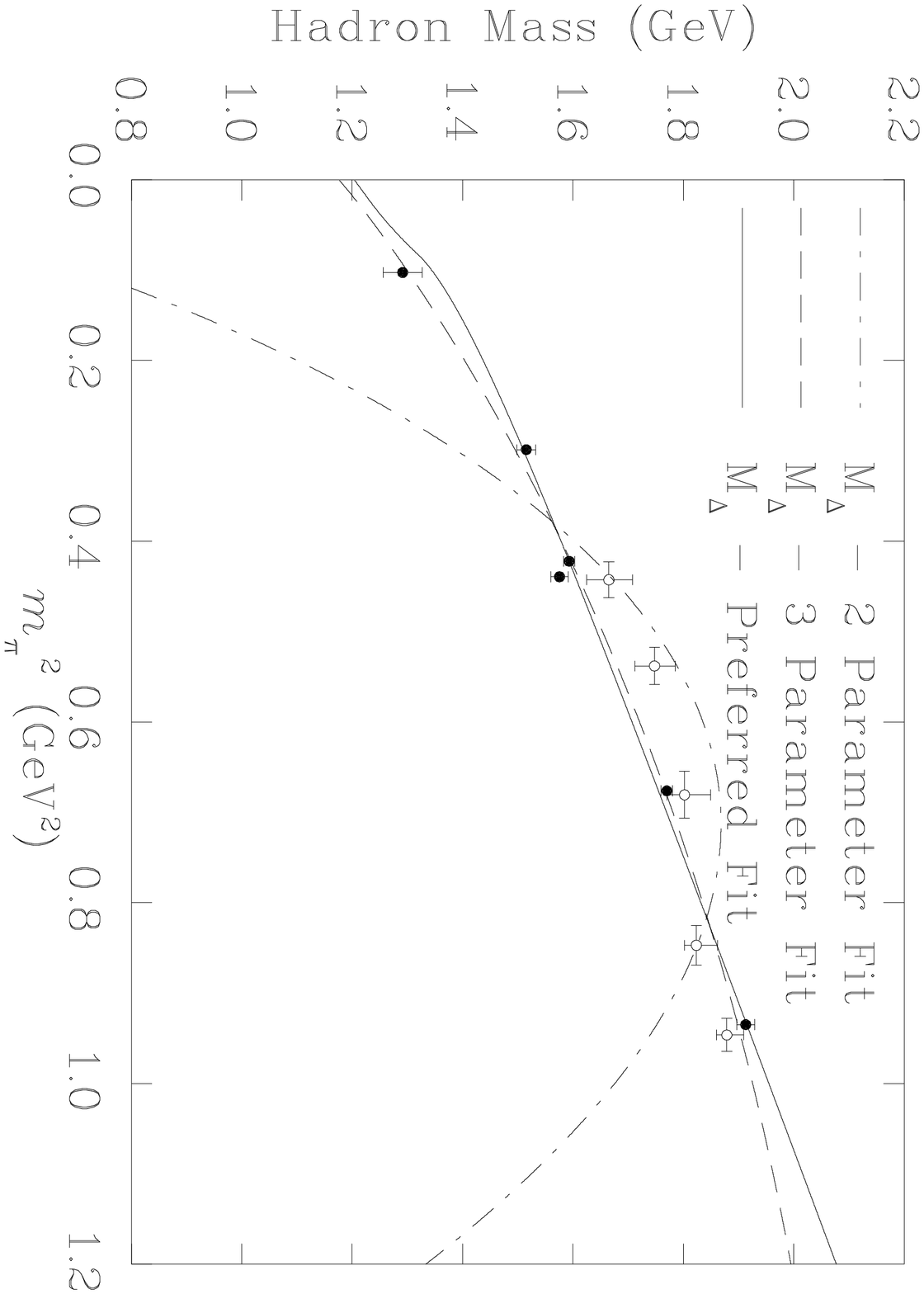,height=12cm}}
\caption{A comparison between phenomenological fitting functions for
the mass of the $\Delta$.  The two parameter fit corresponds to using
Eq. \ref{Naive-eqn} with $\gamma$ set equal to
the value known from $\chi$PT.  The three
parameter fit corresponds to letting $\gamma $
vary as an unconstrained fit parameter.  The solid line is the fit for
the functional form of (\ref{D-form-eqn}), fit (d) of Table
\ref{Lat-Vary-Points-D-table}.
\label{3fits-D-fig}}}
\end{figure}

It is common practice in the lattice community to use a polynomial
expansion for the mass dependence of hadron masses.  Motivated by
$\chi$PT the lowest odd power of $m_\pi$ allowed is $m_\pi^3$:
\begin{equation}
\label{Naive-eqn}
M_{N}=\alpha +\beta m_{\pi }^{2}+\gamma m_{\pi }^{3}
\end{equation}
The results of such fits are shown in Figs. \ref{3fits-N-fig} and
\ref{3fits-D-fig} for $N$ and $\Delta$ respectively.  The
corresponding parameters are reported in table \ref{3fits-params-tab}.
As can be seen in table \ref{3fits-params-tab}, the coefficient of the
$m_{\pi }^{3}$ term, which is the leading non-analytic term
in the quark mass, disagrees with the coefficient known from 
$\chi$PT by almost an order of magnitude. This clearly indicates
the failings of such a simple fitting procedure.  We recommend that
future fitting and extrapolation procedures should be based on
Eqs. (\ref{N-form-eqn}) and (\ref{D-form-eqn}), which are consistent
with $\chi$PT and the heavy quark limit.

\begin{table}[t]
{\centering \begin{tabular}{ccccc}
Fit &
$\alpha $&
$\beta $&
$\Lambda $&
$M_{\Delta}$\\
&(GeV) &(GeV${}^{-1}$) &(GeV) &(GeV) \\
\hline 
(a)&
1.64&
0.414&
0.683&
1.042\\
(b)&
1.37&
0.587&
0.419&
1.240\\
(c)&
1.54&
0.475&
0.616&
1.095\\
(d)&
1.36&
0.602&
0.419&
1.230\\
\end{tabular}\par}
\caption{Parameters for the fits shown in Fig.\
\ref{Lat-Vary-Points-D-fig}.  Parameter sets (a) and (b) are obtained
by excluding the lowest data point from the fit, while (c) and (d)
include it.  Parameter sets (a) and (c) are fits with 3 parameters,
sets (b) and (d) are fits with $\Lambda$ fixed to
the phenomenologically preferred value.
\label{Lat-Vary-Points-D-table}}
\end{table}

\begin{table}[b]
{\centering \begin{tabular}{ccccccccc}
\multicolumn{1}{c}{}&
\multicolumn{4}{c}{$N$}&
\multicolumn{4}{c}{$\Delta$ }\\
Fit &
$\alpha $&
$\beta $&
$\gamma$ or $\Lambda $&
$M_{N}$&
$\alpha $&
$\beta $&
$\gamma$ or $\Lambda $&
$M_{\Delta}$\\
 &(GeV) &(GeV${}^{-1}$) &(GeV${}^{-2}$) or (GeV) &(GeV) 
 &(GeV) &(GeV${}^{-1}$) &(GeV${}^{-2}$) or (GeV) &(GeV) \\
\hline 
(a)&
$-$0.128&
7.38&
$-$5.60&
$-$0.001&
0.182&
7.09&
$-$5.60&
0.304\\
(b)&
0.912&
1.69&
$-$0.761&
0.943&
1.18&
1.45&
$-$0.703&
1.202\\
(c)&
1.15&
0.736&
0.455&
1.003&
1.37&
0.602&
0.419&
1.227
\end{tabular}\par}
\caption{Parameter sets for the fits shown in Figs. \ref{3fits-N-fig}
and \ref{3fits-D-fig}.  Set (a) is for the 2 parameter fit of
(\protect\ref{Naive-eqn}) with $\gamma$ from $\chi$PT, (b) for the 3
parameter fit of (\protect\ref{Naive-eqn}), and (c) for the preferred
functional form.
\label{3fits-params-tab}}
\end{table}

\section{Summary}
\label{discussion-sec}

In the quest to connect lattice measurements with the physical regime,
we have explored the quark mass dependence of the $N$ and $\Delta$
baryon masses using arguments based on analyticity and heavy quark
limits. In the region where $m_\pi$ is larger than 500 MeV, the lattice
data can be reasonably well described by the simple form, $\alpha + 
\beta m_\pi^2$, which is linear in the quark mass. The additional
curvature associated with chiral corrections only appears below this
region. This can be understood quite naturally within 
chiral quark models, like the cloudy bag, which lead to a
cut-off on high momentum virtual pions, thus suppressing the self-energy
diagrams quite effectively as $m_\pi^2$ increases. The pionic
self-energy diagrams which we consider are unique in that only these
diagrams give
rise to the leading non-analytic behaviour which yields a rapid
variation of baryon masses in the chiral limit. Loops involving heavier
mesons or baryons cannot give rise to such a rapid variation. 

Based on these considerations, we have determined a 
method to access quark masses beyond the
regime of chiral perturbation theory. This method reproduces the
leading non-analytic behavior of $\chi$PT and accounts for the
internal structure for the baryon under investigation.  We find that
the predictions of the CBM, and two flavor, dynamical fermion lattice
QCD results, are succinctly described by the formulae of equations
(\ref{N-form-eqn}) and (\ref{D-form-eqn}) with terms defined in
(\ref{NN-Form-eqn}) through (\ref{low-DN-form}).  We believe that
equations (\ref{NN-Form-eqn}) -- (\ref{low-DN-form})
are the simplest one can write down which involve a
single parameter, yet incorporate the essential constraints of chiral
symmetry and the heavy quark limit.

Firm conclusions concerning agreement between the extrapolated lattice
results and experiment cannot be made until the systematic errors can
be reduced below the current level of 10\% and accurate measurements
are made at $m_\pi \sim 300$ MeV or lower.  The significance of
nonlinear behavior in extrapolating nucleon and $\Delta$ masses as a
function of $m_{\pi }^{2}$ to the chiral regime has been evaluated.
We find that the leading non-analytic term of the chiral expansion
dominates from the chiral limit up to the branch point at $m_{\pi
}=\Delta M$. The curvature around $m_{\pi }=\Delta M$, neglected in
previous extrapolations of the lattice data, leads to shifts in the
extrapolated masses of the same order as the departure of lattice
estimates from experimental measurements.

\section*{Acknowledgments}

We would like to thank Pierre Guichon and Tom Cohen for helpful
discussions. This work was supported
in part by the Australian Research Council.


\end{document}